# Searching for cosmological variation of the proton-to-electron mass ratio using a single $H_2$ system

T. D. Le[1,2]

[1] Division of Applied Physics, Dong Nai Technology University, Bien Hoa City, Vietnam
[2] Faculty of Engineering, Dong Nai Technology University, Bien Hoa City, Vietnam
Email: leducthong@dntu.edu.vn

**Abstract.** *Exploring physics beyond General Relativity and the Standard Model of Particle Physics involves investigating spacetime variations in natural constants. This study employs an $H_2$-single of QSO 0347-383 observational spectrum to propose a unique approach for detecting potential changes in the proton-to-electron mass ratio. By comparing the ratio from observational and laboratory data in the Lyman-Alpha transition line, we derive a cosmological variation of $\Delta\mu/\mu = (0.120 \pm 0.144) \times 10^{-8}$ at $z_{abs} = 3.025$. This approach not only advances fundamental physics understanding but also introduces innovative techniques for analyzing high-redshift QSO systems.*

1. **Introduction.**

The Standard Model of Particle Physics (SMPP) and the theory of General Relativity (GR) underlie our present understanding of the universe. Despite their remarkable success, these theories are anticipated to eventually fail at some points. Various scenarios involving unification (Hojjati et al., 2016), higher-dimensional theories (Levshakov et al., 2016), and models concerning dark matter (Martins, 2017) and dark energy (Thompson, 2017; Kostelecky et al., 2003; Uzan, 2011) suggest potential deviations from the Einstein Equivalence Principle (EEP). These theoretical frameworks have led to extensions of physics where constants transition into dynamic phenomena, challenging the EEP.

Investigating spacetime variations in natural constants, such as the fine structure constant, the proton-to-electron mass ratio, and the gravitational constant, serves as a means to test the EEP. Atomic clock experiments have set stringent limits on linear changes in these constants, often at the level of a few parts per million (ppm) (Liberati, 2013; Dzuba et al., 1999; Webb et al., 1999; 2011). Transition wavelengths offer a grat way to detect variations in the fine structure constant ($\alpha$) and the proton-to-electron mass ratio ($\mu$). Atomic clock comparisons provide their temporal derivatives (Blatt et al., 2008; Rosenband et al., 2008), while studies of high-redshift celestial objects explore cosmic-scale changes.

Analysis of high-redshift QSO absorption spectra has indicated potential variations in µ across the universe at a few ppm level (Reinhold et al., 2006). Utilizing microwave and mm-wave molecular transitions in these systems enhances sensitivity to constant variations compared to optical transitions, yielding results for Δμ/μ within the range of 8 to 0.8 ppm (Levshakov et al., 2012; Weiβ et al., 2012; Kanekar et al., 2010; Rahmani et al., 2012). Recent investigations into methanol absorption lines at high redshifts reported a 0.1 ppm change in Δμ/μ. Given the ubiquity of single-molecule lines like hydrogen, probing μ-variations at high redshifts proves advantageous. Additionally, the far-infrared fine structure of $C_I$, $C_{II}$, and CO have been employed to constrain cosmic variations in μ (Levshakov et al., 2020 and references therein). Notably, a study examined nine absorption systems with redshifts ranging from 2.05 to 4.22, yielding a ~5 ppm result at a 3σ level (Ubachs et al., 2016).

The study further employed the μ-dependent state of wavelength transition shifts via the $H_2$-technique (Salumbides et al., 2012). Other works combined 21cm lines of $H_I$ and UV metal absorption lines to assess the temporal variations of μ, with OH microwave line measurements providing combined changes for $\alpha^2 g_p \mu$ ($g_p$ denotes the proton gyromagnetic ratio) (Noterdaeme et al., 2009; 2010). CO, abundant in our galaxies, has also contributed to μ-variation analyses. Investigations involving a mixture of HI and CO in six absorbing systems with high-quality QSO spectra yielded estimates for the cosmic μ-variation (Srianand et al., 2010). In contrast, studies using CH lines in the Milky Way nearly constrained the cosmological μ-variation to a level of Δμ/μ = (-0.07±0.22) ppm (Truppe et al., 2013). Currently, the most robust limit on cosmological μ variation stands at ppm, based on Ni V lines detected in strong gravitational fields from the white dwarf spectrum of G191-B2B (Le., 2019; 2020).

In this study, we focus on the QSO 0347-383 system at a redshift, $z_{abs} = 3.025$, aiming to estimate the cosmological deviation of the proton-to-electron mass ratio. The choice of this high-quality quasar is driven by the objective of constraining Δμ/μ over cosmic timescales. Utilizing a combination of single observational systems and laboratory data, this study provides a significantly improved accuracy determination compared to previous estimates.

## 2. $H_2$ system constraints on Δμ/μ

Astrophysical techniques offer a useful tool for investigating changes in fundamental constants of nature on cosmic timescales. In the context of unification scenarios, constants such as the fine-structure constant and the proton-to-electron mass ratio could exhibit varying values, as indicated by measurements from QSO systems. This exploration of constants involves comparing observed wavelengths from QSO systems with their corresponding laboratory values. For the fine-structure constant's variation, general relativistic effects associated with redshift ($z$) play a central role, involving couplings with scalar and other fields. This is expressed through the loss of energy by a photon($E$), given by $z = -\Delta E/E$. The resulting fractional change in energy leads to a corresponding fractional change in wavelength measurements, $-\Delta E/E = \Delta\lambda/\lambda \sim \Delta\alpha/\alpha$. Spatial or temporal variations in dimensionless physical constants are key elements for unifying gravitational and electromagnetic forces within Grand Unified Theories (GUTs). This unification assumes equivalent Yukawa couplings and is characterized by dimensional transmutations at a weak scale. The framework employs a driven dilaton-type to establish all couplings. This approach relates fine-structure constant variation to the Quantum Chromodynamics (QCD) scale ($\Lambda_{QCD}$) through $\Lambda_{QCD}/\Lambda_{QCD} = R\Delta\alpha/\alpha$, where $R$ is derived from GUTs. By considering model-independence at low energies, the $R$-value can be determined through the relationship $\alpha(M_{GUT}) = \alpha_s(M_{GUT})$. Modifications in Yukawa coupling ($h$) impact the Higgs VEV (Vacuum Expectation Value-$v$) at the Planck mass scale of GUTs. Consequently, dimensional transmutation can deduce $v = M_{Planck}\exp(-(8\pi^{2c})/h^2)$ and $\Delta v/v = 16\pi^2 c(\Delta h/h) = S\ (\Delta h/h)$. Specifically, $S \equiv d\ln v/d\ln h$, and $\Delta h/h = (1/2)\Delta\alpha/\alpha$ ($c \simeq h \simeq 1$). This provides the foundation for testing the variation of the proton-to-electron mass ratio within the framework of unification scenarios (Coc et al., 2007; Ferreira et al., 2014; Le., 2019; 2020; 2021a):

$$\frac{\Delta\mu}{\mu} = [0.8R - 0.3(1 + S)]\frac{\Delta\alpha}{\alpha}$$

It's important to highlight that the values of $R$ and $S$ are determined from astronomical data and exhibit variations across different models (Coc et al., 2007; Ferreira et al., 2014; Le., 2019; 2020; 2021). The present predictions within unification scenarios depend on specific model choices for these parameters. Consequently, the parameters $R = 273 \pm 86$ and $S = 630 \pm 230$ (Le., 2019; 2020; 2021) were selected as the best-fit for our purposes, resulting in significant outcomes above the mentioned values. However, the precise selection of $R$ and $S$ in the context of unified scenarios remains uncertain. To determine a potential cosmological deviation of $\mu$, we utilize these parameter values to compare individual lines from QSO 0347-383 with their associated laboratory. The QSO 0347-383 spectra were acquired using the European Southern Observatory's Ultraviolet and Visible Echelle Spectrograph on the Very Large Telescope (Wend and Molaro, 2011). These spectra provide advantageous qualities with $R = 53000$, $S/N = 30 - 70$, accuracies ranging from 0.2 to 1 ppm for the present analysis. Our analysis reveals that each hydrogen line maintains a uniform and consistent shape. Additionally, we assume uniform velocities for $H_2$. Gaussian line-fitting profiles are employed during the analysis procedure to determine linewidths and central velocities. Single Gaussian fits individual character velocities, while multiple Gaussians accommodate diverse characters. Each character component is denoted as ($N$, $z_{abs}$, $b = \sqrt{2}\,\sigma$), where $N$ represents column density, $z_{abs}$ indicates absorption redshift, $b = \sqrt{2}\,\sigma$ signifies Doppler linewidth, and $\sigma$ denotes root-mean-square (RMS). In our approach, $H_2$ simulated lines were initially generated to estimate $\Delta\alpha/\alpha$ values. The values were subsequently determined using fitting parameters ($\Delta\alpha/\alpha, R, S$). For the given input data, the fitting process only involved ($\chi^2$) and minimal ($\chi^2_{min}$) to derive $\Delta\mu/\mu$ with a reduced fitting $\chi^2 \simeq 1$. The best-fitting value of $\Delta\mu/\mu$ is determined by one-sigma error within standard quality appropriate statistics, signifying the use of to predict $\Delta\mu/\mu$-changes. Subsequently, for estimating errors, the maximum rate changes in $\Delta\mu/\mu$ are based on $\Delta\chi^2 = 1$. The $\chi^2$-minimum is determined for fitting each line, leading to the expected outcome of potential cosmic effects on the proton-to-electron mass ratio, $\Delta\mu/\mu$, as presented in **Table 1**. $\sigma^2_{tot} = \sigma^2_{\Delta\mu/\mu} + \sigma^2_{sys}$ is employed to assess total errors. The redshift-dependent variation of toward QSO 0347-383 is depicted in **Figure 1**.

**Table 1:** The value of $\Delta\mu/\mu = (0.120 \pm 0.144) \times 10^{-8}$ is derived from the weighted average of all $H_2$ lines.

| Line ID | $\lambda_{Obser}$ (Å) | $\lambda_{Lab}$ (Å) | z | $\Delta\alpha/\alpha(10^{-8})$ | $\Delta\mu/\mu(10^{-8})$ | $\sigma_{\Delta\mu/\mu}(10^{-8})$ |
|---|---|---|---|---|---|---|
| L14R1 | 3811.5038 | 946.9804 | 3.0249025 | 0.1073 | 0.49996 | 0.5999 |
| W3Q1 | 3813.2825 | 947.4219 | 3.0249043 | 0.1082 | 0.4196 | 0.5036 |
| W3P3 | 3830.3795 | 951.6719 | 3.024895 | 0.1455 | 0.5875 | 0.7050 |
| L13R1 | 3844.0442 | 955.0658 | 3.0248999 | -0.2218 | 0.2105 | 0.2526 |
| L13P1 | 3846.6271 | 955.7083 | 3.0248966 | 0.0658 | 0.1237 | 0.1484 |
| W2Q1 | 3888.4352 | 966.0961 | 3.0248948 | 0.1574 | 0.2559 | 0.3071 |
| W2Q2 | 3893.2050 | 967.2811 | 3.0248951 | 0.2056 | 0.5173 | 0.6208 |
| L12R3 | 3894.7939 | 967.6770 | 3.0248904 | 0.1653 | 0.2641 | 0.3169 |
| W2Q3 | 3900.3288 | 969.0492 | 3.0249028 | -0.1957 | 0.3910 | 0.4692 |
| L10R1 | 3952.7477 | 982.0742 | 3.0248972 | 0.0112 | 0.2486 | 0.2984 |
| L10P1 | 3955.8160 | 982.8353 | 3.0249022 | 0.1949 | 0.4674 | 0.5609 |
| L10R3 | 3968.3977 | 985.9628 | 3.0248960 | -0.1496 | 0.3415 | 0.4099 |
| L10P3 | 3975.6657 | 987.7688 | 3.0248950 | -0.0213 | 0.1993 | 0.2391 |
| W1Q2 | 3976.4877 | 987.9745 | 3.0248890 | 0.0133 | 0.1204 | 0.1445 |
| L9R1 | 3992.7546 | 992.0164 | 3.0248877 | -0.0326 | 0.1934 | 0.2321 |
| L9P1 | 3995.9594 | 992.8096 | 3.0249000 | 0.2037 | 0.6210 | 0.7452 |
| L8R1 | 4034.7699 | 1002.4521 | 3.0249004 | 0.2055 | 0.5331 | 0.6397 |
| L8P3 | 4058.6575 | 1008.3860 | 3.0249046 | 0.2449 | 0.2647 | 0.3176 |
| W0R2 | 4061.2132 | 1009.0249 | 3.024889 | -0.1014 | 0.5395 | 0.6474 |
| L7P3 | 4103.3836 | 1019.5022 | 3.0248894 | 0.0094 | 0.1415 | 0.1699 |
| L6R3 | 4141.5640 | 1028.9866 | 3.024896 | -0.2418 | 0.2847 | 0.3416 |
| L6P3 | 4150.4349 | 1031.1926 | 3.0248882 | -0.1325 | 0.1921 | 0.2305 |
| L5R1 | 4174.4204 | 1037.1498 | 3.0248963 | 0.0353 | 0.1880 | 0.2257 |
| L5R2 | 4180.6152 | 1038.6903 | 3.024891 | -0.1914 | 0.2917 | 0.3501 |
| L5R3 | 4190.5690 | 1041.1588 | 3.0249086 | -0.0383 | 0.1277 | 0.1532 |
| L4R1 | 4225.9822 | 1049.9597 | 3.0248994 | 0.1573 | 0.2239 | 0.2687 |
| L4P1 | 4230.2974 | 1051.0325 | 3.0248969 | -0.2398 | 0.4084 | 0.4901 |
| L4R3 | 4242.1531 | 1053.9761 | 3.0249045 | -0.2238 | 0.3238 | 0.3886 |
| L4P3 | 4252.1911 | 1056.4714 | 3.0248994 | -0.0185 | 0.1959 | 0.2351 |
| L3R1 | 4280.3234 | 1063.4601 | 3.0249027 | -0.1292 | 0.2924 | 0.3509 |
| L3P1 | 4284.9349 | 1064.6054 | 3.0249043 | 0.2214 | 0.5214 | 0.6257 |
| L3R3 | 4296.4822 | 1067.4786 | 3.0248884 | -0.1590 | 0.5590 | 0.6709 |
| L3P3 | 4307.2114 | 1070.1409 | 3.0249012 | 0.0342 | 0.1742 | 0.2091 |
| L2P3 | 4365.2399 | 946.9804 | 3.0248937 | -0.0853 | 0.1853 | 0.2224 |
| L1R1 | 4398.1291 | 947.4219 | 3.0248913 | -0.1355 | 0.5355 | 0.6426 |
| L1P1 | 4403.4456 | 951.6719 | 3.0248961 | -0.1263 | 0.2263 | 0.2716 |

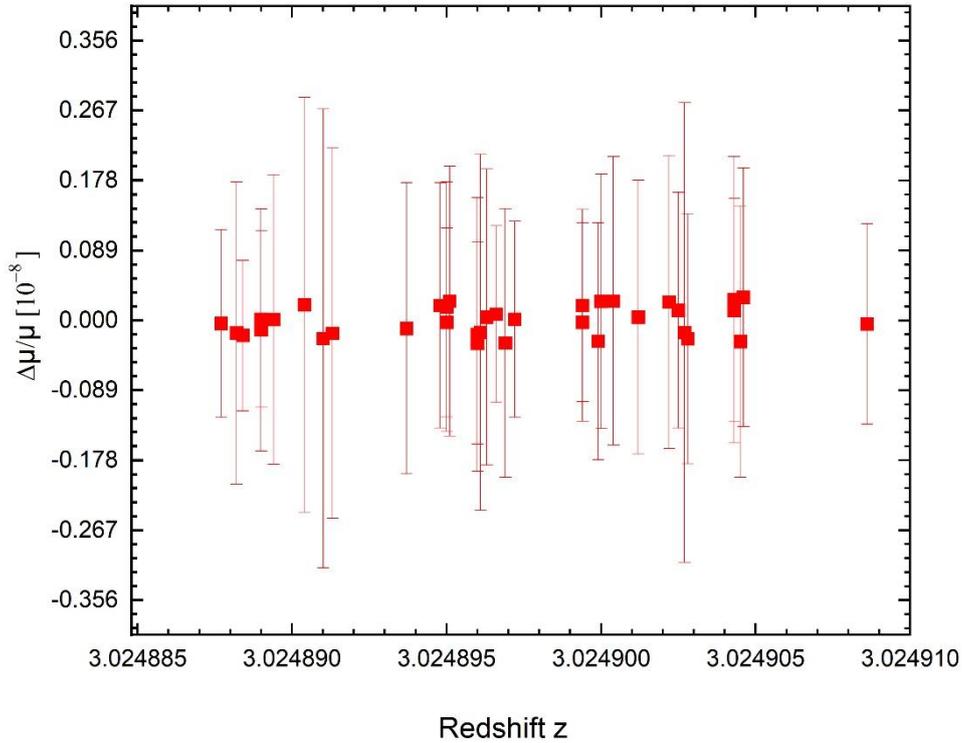

**Figure 1.** A plot for Δμ/μ versus redshifts

The single $H_2$ lines offer a means to test limitations on natural constants such as $\alpha$ and μ. However, discrepancies arise when different techniques are applied to the same observational data, revealing challenges in control and understanding. These variations derive from uncertainties in wavelength laboratories and their associated analysis methodologies, leading to different conclusions (Ubachs et al., 2016; Calmet and Fritsch, 2002; Tzanavarris et al., 2005; Salumbides et al., 2012; Kostelecky et al., 2003; King et al., 2012).

In this context, our present analysis uses a nonlinear least-squares approach, applying observable spectra with uncertainties ranging from 0.2 to 1 ppm and updated laboratory wavelengths. This technique effectively quantifies Δμ/μ -errors with remarkable precision through Gaussian distribution (Le., 2019; 2020; 2021a, b, c). Our study aims to demonstrate that a single $H_2$ system can set constraints on potential variations in physical constants, such as the fine-structure constant and the proton-to-electron mass ratio.

It's crucial to emphasize that the current study is particularly relevant for determining variations in the proton-to-electron mass ratio within high-redshift systems. As a result, we establish a novel limit of $\Delta\mu/\mu = (0.120 \pm 0.144) \times 10^{-8}$ at the absorption redshift $z_{abs} = 3.025$, basing on the analysis of $H_2$ lines.

3. **Discussions**

Exploration of changes in fundamental constants across cosmic spacetime will generate new cosmological models and mysterious phenomena beyond the SMPP. Observational data provides as a powerful tool to proble these phenomena. Recent investigations have unveiled a series of unified hypotheses concerning spacetime variations in fundamental constants, particularly $\alpha$ and $\mu$. Hence, studying the cosmological evolution of these constants plays an importantl role in advancing unification theories.

Recently, astronomical measurements have yielded important outcomes. For instance, measurements involving the 6-bearing damped Lyman systems within the redshift range of $2 \leq z \leq 3$ resulted in a finding of $\Delta\mu/\mu \leq 10$ ppm (Levshakov et al., 2002; Malec et al., 2010; Varshalovich and Levshakov, 1993; Wend and Molaro, 2011; Kanekar, 2011). Combining the same data with other molecules, researchers established $\mu/\mu = -(0.35 \pm 0.12)$ ppm and $\Delta\mu/\mu = (0.10 \pm 0.47)$ ppm with $z \leq 1.0$, respectively, along with differential conclusions drawn from studying a single system toward Q 0528-250, which resulted in $\Delta\mu/\mu = (0.3 \pm 3.7)$ ppm (Rahmani et al., 2012; 2013). Additionally, analysis of ammonia toward the quasar PKS-1830-211 at a redshift of $z = 0.89$ yielded a $\Delta\mu/\mu = (0.00 \pm 0.10)$ ppm. However, this study only focused on the low-redshift system $z \leq 1.0$ applying the $NH_3$ and $CH_3OH$ lines (Levshakov et al., 2011). The investigation of the quasar J1337+3152 produced a $\Delta\mu/\mu = (-1.7 \pm 1.7)$ ppm measurement at $z \sim 3.17$, while another study set a lower limit of $\Delta\mu/\mu = (0.0 \pm 1.5)$ ppm with $z_{abs} \sim 1.3$ utilizing a selection of four 21-cm absorption systems (Levshakov et al., 2010a, b). Consequently, recent research has largely limited the cosmological variation of the proton-to-electron mass ratio to the $\sim 0.0028$ ppm level (Levshakov et al., 2011).

Proposing an effective approach for increasing this limit associates realizing on single $H_2$ systems, using high-resolution spectra from existing optical instruments. Single $H_2$ systems, given their normality across the universe, prove more crucial than other multiplet lines or laboratory setups. Throughout our study, we identified that the width-separation ratio significantly contributes to major errors, which could be a small separation related to observational and laboratory wavelengths.. This source of uncertainty represents the main error and varies between absorption systems. When a large number of observations are used, these uncertainties lead to average out. Our estimations of potential error sources in the cosmic variation in µ have greatly advanced due to these uncertainties (0.2-1 ppm), which occur due to the relationship between observational and laboratory wavelength calibrations. Moreover, the influence of the Doppler shift effect has been identified and quantified through these error sources, which will not affect of our results. A linear mapping between realistic and measurable calibration error separations can yield highly accurate findings with ∆µ/µ-dependence. Consequently, potential error sources may deriver from observations or laboratory-based calculations involving the $\chi^2$-computation.

## 4. Conclusions.

Compact astronomical objects are becoming prominence as testing for dimensionless physics paradigms. In this study, we have indecated the predictive capability of single $H_2$ systems in evaluating the impact of cosmic spacetime variations on fundamental constants, specifically the proton-to-electron mass ratio's correlation with the fine-structure constant and phenomenological parameters ( $R$ , $S$ ). Our investigation establishes an upper limit of ∆µ/µ within this context.

We have demonstrated that the combination of single $H_2$ systems yields constraints on phenomenological parameters that characterize unification models. This exploration probes into how the relationships among physical constants are influenced within the framework of GUT models, where variations in both α and µ are expected to vary. Notably, the limitations set by single lines are better than those set by multiplet lines and atomic clock tests in the context of spacetime variations of these constants.

As a result, we have established an upper limit of ∆µ/µ within this context, offering precision beyond previous studies that relied on multiplet lines and atomic clock experiments to examine spacetime variations in µ. Moreover, this achievement is attributed to spatial/temporal approximations in extra-dimensional contexts at the level of $\Delta\mu/\mu = (0.120 \pm 0.144) \times 10^{-8}$ .

The study of spatial or temporal variations in natural constants (Murphy et al., 2008; Uzan, 2011; Bagdonaite et al., 2013; Liberati, 2013; Will, 2014; Martins, 2017; Le., 2019; 2020; 2021a, b, c) generally contributes to novel theoretical physics extending beyond current observational models. The method employed to probe these variations, such as cosmic changes in µ, provides potential for determination on the incomplete nature of the Einstein Equivalence Principle (EEP). The pursuit of more precise measurements of these constants from astrophysical sources spanning diverse redshift ranges should guide future research endeavors. In any scenario, this study broadens the figure of possibilities for identifying features within Grand Unified Theories (GUTs).

### 5. Data availability

The data used to support the findings of the present study are listed in **Table 1**.

**Declaration of Competing Interest**

The author has no conflict of interest to declare.

**Funding**

This research has not been received external funding.